\documentstyle[11pt,newpasp,twoside,epsf]{article}
\def\edcomment#1{\iffalse\marginpar{\raggedright\sl#1\/}\else\relax\fi}
\reversemarginpar

\begin{document}
\title{The Onsala blind 6.7 GHz survey of the galactic plane: new methanol masers in the northern hemisphere.}
 \author{Michele R. Pestalozzi$^1$, Vincent Minier$^1$, Roy Booth$^1$, John Conway$^1$}
\affil{$^1$ Onsala Space Observatory, S-439 92 Onsala, Sweden\\}

\begin{abstract}
We review the state of the Onsala blind survey of the galactic plane, searching
for new 6.7 GHz methanol masers. We also describe preliminary
results of millimeter follow-up observations of the new detections and high
resolution observations using the EVN. We conclude that blind surveys are
important to complement targeted searches done until
now and give the possibility to detect new classes of objects.
\vskip -1cm
\end{abstract}

\section{Introduction}

Methanol masers are often detected in regions with typical signposts of massive star
formation, such as H$_2$O and OH masers, and some of them are detected towards
IRAS sources satisfying the Wood \& Churchwell (1989) criterium for UCHII
regions. Nevertheless high resolution VLBI observations show that such masers
are not necessarily cospatial with IRAS
UCHII regions (Minier, Conway, \& Booth 2001), which leads us to suppose that the maser activity can appear at
a very early stage in the formation of a star, before the formation of a UCHII
region. This fact increase the importance of blind surveys for 6.7 GHz masers of the whole
galactic plane.

We are using the Onsala 25m telescope to conduct such a survey ($\mid b \mid
\le 0.55^{o}$, $35^{o} \le l \le 220^{o}$), searching for the
powerful 6.7 GHz maser line of methanol. Since late 1998 we have surveyed
about 45 deg$^2$, and have made 11 detections (see Table 1). An important
aspect of this survey lies in the fact that methanol maser is the only maser
emission which has been searched for in this way in the northern hemisphere. Our unbiased survey should be regarded as complementary to the work carried out towards the Galactic center from Australia (Caswell 1996a+b, Ellingsen 1996).

The newly upgraded Onsala 25m telescope is equipped with a three stage
cryogenically cooled High Electron Mobility Transistor (HEMT) amplifier
mounted at the secondary focus. The backend consists of a 1600 channel autocorrelator
fed with a 6.4 MHz bandwidth, which guarantees a complete coverage of the
velocity distribution in the Galaxy ($\pm$ 150 km/s). Each square degree of
the galactic plane is surveyed following an equilateral triangular grid, integrating 10 minutes per
position. The system temperature lies between 1200 and 1500 Jy. Our sensitivity
is about 6 Jy/channel at 5 $\sigma$ rms. The positions of 10 of the 11 detections were
accurately determined using the 100m Effelsberg telescope (September 2000, 10
arcsec accuracy) and those positions were used for the follow-up VLBI and millimeter observations.

\section{Results from the blind survey}

Among the 11 detections, some masers have already been detected during other
surveys; this allows us to test our system sensitivity. 5 detections have
apparently been missed by previous pointed surveys. This allows us to make some
conclusions about the selection criteria. For instance, most of the pointed
surveys conducted until now made use of color filtering of the IRAS
Point Source Catalogue (PSC). The reason for that is that class II methanol
masers (such as that at 6.7 GHz) are supposed to appear in the
neighbourhood of a strong infrared source which guarantees the pumping of the
non thermal maser emission. A cross check of the detection rate in previous
pointed surveys shows a relatively low efficiency in terms of number of detections of all
those selection criteria: except for the surveys by Menten et al. (1991) and
Caswell et al. (1995), the average detection rate is about 20\% (see Table 2). 

\begin{table}[!h]
\begin{center}
\caption{Source list of detected 6.7 GHz methanol masers with coordinates in
  J2000, flux density (Jy), LSR velocity and velocity range of the strongest maser
  feature (km/s), as of May 2001. The source marked with a star has been detected in 
 March 2001, its position comes from measurements at the 25m telescope at
  Onsala (30 arcsec accuracy). In the last column references to previous observations are listed: 1) Szymczak et al. 2000, 2) Menten 1991, 3) Caswell et al. 1995.}
\begin{tabular}{lcccccc}
\tableline
Source name & RA (J2000) & Dec (J2000) & S$_{peak}$ & $v_{peak}$ & ${\Delta}v$  & Reference \\
\tableline
G 35.88-0.20$\star$ & 18:57:30.7 & 02:31:45  & 3.0 & 60 & 55, 65 & new det \\
G 40.25-0.19 & 19:05:32.6 & 06:25:38  & 12 & 70 & 69, 71 & 1 \\
G 40.61-0.09 & 19:05:50.2 & 06:47:18  & 4.5 & 27 & 26, 28 & 2 \\
G 41.34-0.14 & 19:07:22.1 & 07:25:12  & 51 & 14 & 13, 16 & new det \\
G 42.07+0.24 & 19:07:20.8 & 08:14:13  & 11 & 12 & 10, 14 & new det \\
G 43.17+0.01 & 19:10:14.3 & 09:06:14  & 15 & 2 & -2, 16 & 2 (W49) \\
G 43.80-0.13 & 19:11:54.3 & 09:35:53  & 30 & 37 & 34, 41 & 2 \\
G 44.05+0.00 & 19:11:54.3 & 09:53:02  & 4.5 & 40 & 38, 42 & new det \\
G 45.07+0.13 & 19:13:22.4 & 10:50:58  & 31 & 57 & 56, 58 & 2 \\
G 45.53+0.13 & 19:14:14.8 & 11:15:16  & 5.5 & 63 & 53, 65 & 3 \\
G 85.40-0.07 & 20:54:15.3 & 44:53:54  & 45 & -30 & -32, -28 & new det \\
\tableline\tableline
\end{tabular}
\end{center}
\end{table}


\begin{table}[!h]
\begin{center}
\caption{Comparison between biased surveys and blind surveys of
  methanol. Notice the low efficiency of the IRAS color selected surveys in
  terms of detections. The particularly low detection rate of the Onsala
  survey can be the result of several factors: from Onsala we do not see the
  galactic center, fewer spiral arms are visible, slight lower sensitivity and
  possible variability of the sources.}
\begin{tabular}{lll}
\\
\tableline
{\bf Biased surveys} & & \\
\tableline
Menten (1991) & 80 / 123 & UCHII, H$_{2}$O, OH maser sites \\
Schutte et al. (1993) & 35 / 235 & IRAS, color selected \\
Caswell et al. (1995) & 245 & OH (208), star formation regions\\
van der Walt et al. (1995) & 31 / 520 & IRAS (color selected UCHII) \\
Walsh et al. (1997) & 201 / 535 & IRAS (color selected) \\
Slysh et al. (1999) & 42 / 429 & IRAS (color selected UCHII)\\
Szymczak et al. (2000) & 182 / 1399 & IRAS (color selected) \\
\multicolumn{3}{l}{{\bf $\to$ Detection rate towards IRAS sources: $\sim$ 20 \%}}\\ 
 & & \\
\tableline
{\bf Blind surveys} & & \\
\tableline
Caswell (1996a,b) & 80 masers in 11 deg$^2$ & $\to$ 7.2 masers / deg$^2$ \\
Ellingsen (1996) & 107 masers in 33 deg$^2$ & $\to$ 3.2 masers / deg$^2$ \\
Onsala (1999-2001) & 11 masers in 45 deg$^2$ & $\to$ $\sim$ 0.25 masers / deg$^2$ \\
\tableline\tableline
\end{tabular}
\end{center}
\end{table}

A comparison between IRAS-targeted and blind surveying in the regions
observed at Onsala shows that, if we had applied a targeted search, we would have
missed between 5 and 7 methanol maser sources, depending on the filtering
applied to the IRAS catalogue (we considered three possiblities, Szymczak et al. 2000, Wood \& Churchwell 1989, Schutte et
al. 1993, ordered by decreasing
number of IRAS sources included). This fact strongly justifies the worth of a blind survey.

Where no IRAS counterpart was found, we searched for Mid-IR sources around
the methanol maser positions listed in the MSX catalogue (Midcourse Space
Experiment). The fact is that for some sources not even a Mid-IR counterpart
is found indicating that the methanol masers must arise in very compact and deeply
embedded sources where only longer wavelengths ($\lambda > 100 \mu m$) can pass through the optically thick dust cocoon. Further observations at 450 and 850 $\mu$m will be hopefully performed in order to clarify the origin of these sources.

\section{Follow-up observations}

The follow-up program of the Onsala survey consists of both a searching for and
mapping of traditional high density tracers ($^{13}$CO, C$^{18}$O, CS,
HCO$^+$, HCN, thermal CH$_{3}$OH, HC$_{3}$CN) in the millimeter band (85-110 GHz)
using the Onsala 20m telescope, and high resolution observations using the
EVN. Because of the spectral simplicity of the new methanol masers (see
e.g. Figure 1 and for comparison Minier et al. in this volume) it is difficult to make conclusions about the shape
of the emitting region. The maps in the millimeter range suggest, on the other
hand, that the methanol masers could be offset from the highest density gas, as
shown in Figure 2.

\vskip 0.3cm


\begin{figure}[!h]
\plotone{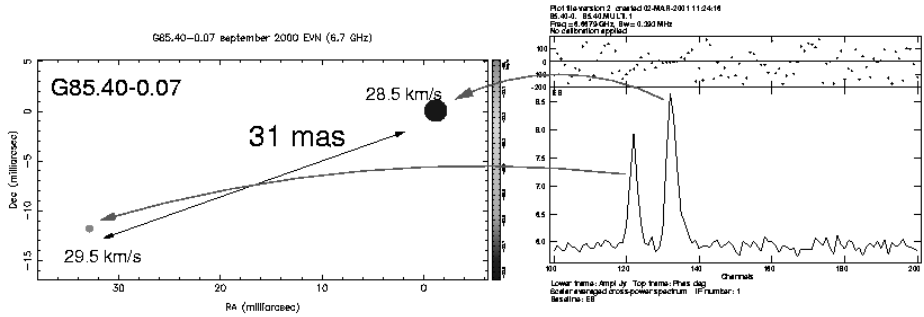}
\caption{EVN cross-spectrum (baseline EB-TR) and relative position map of the source
  G~85.40-0.07. The simplicity of the spectrum does not allow any conlcusion
  about the shape of the emitting region. The distance between the two
  emission points is 160 AU at the estimated kinematical heliocentric distance of
  6 kpc. The velocity dispertion of the maser feature in the spectrum is
  2 km/s.}
\end{figure}
\vskip -0.6cm
\begin{figure}[!h]
\plotone{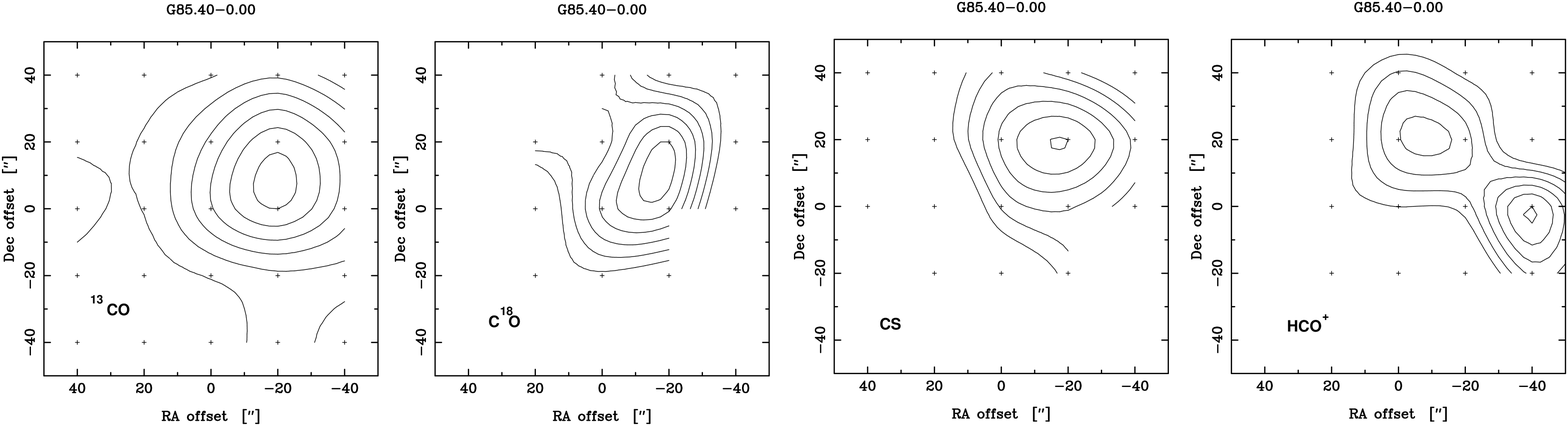}
\caption{mm maps in $^{13}$CO, C$^{18}$O, CS and HCO$^{+}$ around the source
  G~85.40-0.07. The first contour is at 50\% of the peak emission, spacing 10\%. Map spacing is 20 arcsec. The methanol maser is located at the (0,0) position. The position accuracies are 5 arcsec for the maps, 10 arcsec for the methanol maser.}
\end{figure}

\end{document}